\documentclass[pra, twocolumn, superscriptaddress, amsfonts, amsmath,floatfix]{revtex4-1}

\usepackage[english]{babel}
\usepackage{graphicx}
\usepackage{color}
\usepackage{amssymb,amsmath}
\usepackage{xspace}
\usepackage{upgreek}
\usepackage{ulem}
\usepackage{psfrag}
\usepackage{filemod}
\usepackage{siunitx}
\newcommand{\be}{\begin{equation}}
\newcommand{\ee}{\end{equation}}

\newcommand{\rg}{{\bf r}}

\newcommand{\Eg}{{\bf E}}

\newcommand{\Gg}{{\bf G}}

\newcommand{\eg}{{\bf e}}
\newcommand{\jg}{{\bf j}}
\newcommand{\pg}{{\bf p}}

\newcommand{\ug}{\hat{\bf u}}
\newcommand{\xg}{\hat{\bf x}}
\newcommand{\yg}{\hat{\bf y}}

\newcommand{\V}{\cal V}
\newcommand{\QEg}{\tilde{\bf E}}
\newcommand{\Qomega}{\tilde{\omega}}

\begin{document}

\title{Purcell effect with extended sources: The role of the cross density of states}

\author{R. Carminati} 
\altaffiliation{remi.carminati@espci.psl.eu}
\affiliation{Institut Langevin, ESPCI Paris, PSL University, CNRS, 1 rue Jussieu, F-75005 Paris, France}
\author{M. Gurioli}
\altaffiliation{massimo.gurioli@unifi.it}
\affiliation{Physics and Astronomy Department, Universit\`a degli Studi di Firenze, via G. Sansone 1,
50019 Sesto Fiorentino, Italy}

\begin{abstract}
We analyze the change in the spontaneous decay rate, or Purcell effect, of an extended quantum emitter in a structured photonic environment. Based on a simple theory, we show that the cross-density of states is the central quantity driving interferences in the emission process. Using numerical simulations in realistic photonic cavity geometries, we demonstrate that a structured cross-density of states can induce subradiance or superradiance, and change substantially the emission spectrum. Interestingly, the spectral lineshape of the Purcell effect of an extended source cannot be predicted from the sole knowledge of the spectral dependence of the local density of states.
\end{abstract}

\maketitle

\section{Introduction}
In the last two decades, the ability to tailor the spontaneous decay rate of a quantum emitter, by molding the dielectric environment at the nanoscale, has been the driving force of a large variety of studies in cavity quantum electrodynamics~\cite{cavityQED}, semiconductor exciton physics~\cite{Shields2007}, and nanophotonics~\cite{Stockman-review,Novotny-book,Carminati-review}. The engineering of the electromagnetic local density of states (LDOS), in order to tune the spontaneous decay rate by the Purcell effect~\cite{Purcell1946}, has become a major branch of nanophotonics and plasmonics. The standard formulation of the Purcell effect is based on the electric dipole approximation, which requires the electromagnetic field to be uniform over the spatial extent of the emitter. The dipole approximation is expected to breakdown when extended sources are used (for example extended excitons in semiconductors~\cite{Andreani1999}), or when the external field is confined at the nanometer scale, or beyond~\cite{Benz2016}. A theory governing the spontaneous emission by extended excitons has been proposed in Ref.~\cite{Lodahl2012}. The theory predicts large oscillator strengths, exceeding $10^3$ for quantum-dots (QDs) with diameters on the order of $100$~nm. These ideas led to the concept of single-photon superradiance, which has been observed in natural QDs arising in monolayer fluctuation of GaAs quantum wells, with radiative lifetime below $100$~ps.  On a different scheme, Rydberg excitons have been observed in CuO$_2$ bulk samples, with a Bohr radius extending over micrometers~\cite{Kazimierczuk2014,Steinhauer2020}. Extended sources are also at play in microwave cavity quantum electrodynamics with Rydberg atoms~\cite{Haroche2013}, and in electron spectroscopy where the interaction with light depends on non-local excitations~\cite{Abajo2010}. In a slightly different context, understanding and controlling the emission of a set of single-photon sources connected through their electromagnetic environments has become a key issue for the treatment of quantum information in photonics~\cite{Kimble2008,Pryde2019}, or to favor collective emission~\cite{Pustovit2009,Oppel2014}. This has stimulated the development of new theoretical approaches to describe photonic states in ensembles of quantum emitters~\cite{Chang2017,Feist2021}. In the simple case of two (point) quantum emitters, the role of the cross density of states (CDOS), that describes electromagnetic spatial coherence between different points in an arbitrary environment~\cite{Caze2013}, has been put forward~\cite{Canaguier2016,Canaguier2019}. It was shown that the degree of quantum coherence in the emitted light is controlled by a combination of the LDOS and CDOS, the latter describing the coupling between the sources in the emission process.

In the case of extended quantum sources that cannot be reduced to a single point dipole or an ensemble of independent (mutually incoherent) point sources, it has been shown that spontaneous decay results from a non-local interaction with the environment~\cite{Lodahl2012}. Nevertheless, the exact role of the CDOS on the Purcell effect has not been explored in its full generality. In particular, beyond the generation of large effective oscillator strengths, one can expect an engineering of the CDOS to affect the emission substantially, {\it e.g.} by producing subradiance or superradiance, or by changing the emission spectrum. The purpose of this paper is to analyze precisely the Purcell effect for an extended quantum source. Based on simple theoretical arguments, we show that the CDOS is the essential ingredient to understand the change in the spontaneous emission decay rate, beyond the dipole approximation. The theoretical results are highlighted by numerical simulations, in a series of simple examples involving either a single high-$Q$ cavity or coupled cavities with non Lorentzian spectra. Interestingly, the spectral emission lineshape of an extended source covering a nanostructured environment cannot be predicted from the spectral dependence of the LDOS alone.

\section{Spontaneous decay rate in structured environments}

In this section we introduce the expressions of the spontaneous decay rate of point and extended dipole emitters. We assume a weak-coupling regime, in which the population of the excited state exhibits exponential decay, and the transition dipole is not modified by the interaction with the environment. We also restrict the discussion to electric-dipole transitions, and the theory involves the electric part of the density of states of the electromagnetic field. For magnetic-dipole transitions, similar expressions could be easily deduced by using the magnetic contribution to the density of states~\cite{Carminati-review,Joulain2003,Aigouy2014}.

\subsection{Dipole emitter}

In the weak-coupling regime, the spontaneous decay rate $\Gamma$ of a two-level dipole emitter can be calculated from perturbation theory. For an emitter with transition dipole $\pg$, emission frequency $\omega$, and located at the point $\rg_s$, it can  be shown that~\cite{Agarwal1975,Sipe1984}
\begin{equation}
\Gamma(\rg_s,\omega) = \frac{2 \mu_0 \, \omega^2}{\hbar} \, |\pg|^2 \, \mathrm{Im} \left [ \ug \cdot \Gg(\rg_s,\rg_s,\omega) \ug \right ] \, .
\label{eq:Gamma_ImG_general}
\end{equation}
In this expression $\Gg$ is the electric Green's function describing the electromagnetic response of the environment, and $\ug=\pg/|\pg|$ is the unit vector defining the orientation of the transition dipole. We note that the internal dynamics of the emitter, described by the transition dipole $\pg$ that depends on the excited and ground state wavefunctions, and the response of the environment factorize.
In free space, the Green's function $\Gg_0$ is such that $\mathrm{Im} \left [ \ug \cdot \Gg_0(\rg_s,\rg_s,\omega) \ug \right ] = \omega/(6\pi c)$, with $c$ the speed of light in vacuum, and the spontaneous decay rate is $\Gamma_0(\omega) = \omega^3|\pg|^2/(3 \pi \hbar \,\epsilon_0 \, c^3)$. Normalizing, we find that
\begin{equation}
\frac{\Gamma(\rg_s,\omega)}{\Gamma_0(\omega)} = \frac{6 \pi c}{\omega}  \, \operatorname{Im} \left [ \ug \cdot \Gg(\rg_s,\rg_s,\omega) \ug \right ] \ .
\label{eq:generalized_Purcell_Gamma}
\end{equation}
It will prove useful to recall a well-established correspondence between classical and quantum calculations. Considering a classical electric dipole $\pg$ oscillating at frequency $\omega$, the time-averaged power released in the environment is $P(\rg_s,\omega) = (\mu_0 \omega^3/2) |\pg|^2 \mathrm{Im} \left [ \ug \cdot \Gg(\rg_s,\rg_s,\omega) \ug \right ]$. Normalizing by the power $P_0$ radiated in free space, we obtain
\begin{equation}
\frac{P(\rg_s,\omega)}{P_0(\omega)} = \frac{6 \pi c}{\omega}  \, \operatorname{Im} \left [ \ug \cdot \Gg(\rg_s,\rg_s,\omega) \ug \right ] \ .
\label{eq:normalized_classical_power}
\end{equation}
The right-hand side is the same as in Eq.~(\ref{eq:generalized_Purcell_Gamma}), showing that the change in the spontaneous decay rate of a quantum emitter due to the electromagnetic interaction with its environment, and the change in the power emitted by a classical dipole, are identical~\cite{Novotny-book,Carminati-review,Bouchet2019}.

In the particular case of a single mode cavity, the right-hand side in Eqs.~(\ref{eq:generalized_Purcell_Gamma}) and (\ref{eq:normalized_classical_power}) defines the Purcell factor. For an emitter on resonance, it is usually written as $3/(4\pi^2) \lambda_m^3 Q/\V$, with $Q$ the mode quality factor, $\V$ the mode volume and $\lambda_m$ the resonance wavelength~\cite{Purcell1946,Carminati-review}. Equations (\ref{eq:generalized_Purcell_Gamma}) and (\ref{eq:normalized_classical_power}) generalize the Purcell effect to a more complex environment, and the right-hand side can be seen as a generalized Purcell factor. This factor actually describes the change in the projected LDOS
\begin{equation}
\rho(\rg_s,\omega) = \frac{2 \omega}{\pi \, c^2} \, \mathrm{Im} \left [ \ug \cdot \Gg(\rg_s,\rg_s,\omega) \ug \right ] \, ,
\label{eq:partial_LDOS_def}
\end{equation}
which measures the coupling strength of a unit dipole $\ug$ to the electromagnetic environment. The LDOS actually counts the relative contribution of the electromagnetic modes at position $\rg$ and frequency $\omega$~\cite{Novotny-book,Carminati-review,Sauvan2013}. In terms of the LDOS we find that
\begin{equation}
\frac{\Gamma(\rg_s,\omega)}{\Gamma_0(\omega)} = \frac{P(\rg_s,\omega)}{P_0(\omega)} = \frac{ \rho(\rg_s,\omega) }{\rho_0(\omega) } \, ,
\label{eq:Purcell_dipole}
\end{equation}
where $\rho_0(\omega) = \omega^2/(3\pi^2 c^3)$ is the projected LDOS in free space.

\subsection{Purcell effect for an extended source}

A full quantum theory of spontaneous emission by an extended exciton has been developed in Ref.\cite{Lodahl2012}, emphasizing the role of the internal dynamics of the source. In the present work, we rather focus on the role of the environment, and take advantage of the equivalence of the classical and quantum descriptions to build the expression of the decay rate starting from the expression of the emitted power. 
For a monochromatic classical source with current density $\jg(\rg)$ confined within a volume~$V$, and oscillating at frequency $\omega$, the emitted power can be written as
\begin{equation}
P = -\frac{1}{2} \mathrm{Re} \int_V \jg^*(\rg) \cdot \Eg(\rg) \, d^3r \, ,
\end{equation}
where the superscript $*$ stands for complex conjugate. In terms of the Green's function $\Gg$ this becomes
\begin{equation}
P = \frac{\mu_0 \omega}{2} \int \mathrm{Im} \left [ \jg^*(\rg) \cdot\Gg(\rg,\rg^\prime,\omega) \, \jg(\rg^\prime) \right ] \, d^3r \, d^3r^\prime \, ,
\end{equation}
which, by making use of the reciprocity relation $\Gg(\rg,\rg^\prime,\omega) = \Gg^T(\rg^\prime,\rg,\omega)$, can be rewritten as
\begin{equation}
P = \frac{\mu_0 \omega}{2} \int \jg^*(\rg) \cdot \mathrm{Im}\Gg(\rg,\rg^\prime,\omega) \, \jg(\rg^\prime) \, d^3r \, d^3r^\prime \, .
\end{equation}
Normalizing by the power emitted in free space, we find that
\begin{equation}
\frac{P}{P_0} = \frac{\int \jg^*(\rg) \cdot \mathrm{Im}\Gg(\rg,\rg^\prime,\omega) \, \jg(\rg^\prime) \, d^3r \, d^3r^\prime}{\int \jg^*(\rg) \cdot \mathrm{Im}\Gg_0(\rg,\rg^\prime,\omega) \, \jg(\rg^\prime) \, d^3r \, d^3r^\prime} \, .
\label{eq:Purcell_extended_interm}
\end{equation}
In the classical picture, the current density corresponding to an extended exciton can be taken of the form
\begin{equation}
\jg(\rg) = -i\omega \ug \, \eta(\rg,\rg_s) \, ,
\end{equation}
where $\eta(\rg,\rg_s)$ is the volume density of transition dipole at the position $\rg$, $\ug$ gives the local orientation of the transition dipole and $\rg_s$ defines the position of the source (the reference point $\rg_s$ can be chosen arbitrarily, for example as the center of the source spatial distribution). This model allows to take into account a possible heterogeneity of the distribution of elementary dipoles, in amplitude and orientation, inside the extended emitter. {Also note that a single extended exciton corresponds to spatially coherent source, that can be seen as a continuous distribution of mutually coherent elementary transition dipoles.} Inserting this expression into Eq.~(\ref{eq:Purcell_extended_interm}), and invoking again the classical to quantum correspondence, we obtain
\begin{equation}
\frac{\Gamma(\rg_s,\omega)}{\Gamma_0(\omega)} = \frac{\int \eta^*(\rg,\rg_s) \, \eta(\rg^\prime,\rg_s) \, \mathrm{Im} [\ug \cdot \Gg(\rg,\rg^\prime,\omega)  \ug^\prime] \, d^3r \, d^3r^\prime}{\int \eta^*(\rg,\rg_s) \, \eta(\rg^\prime,\rg_s) \, \mathrm{Im} [\ug \cdot \Gg_0(\rg,\rg^\prime,\omega)  \ug^\prime] \, d^3r \, d^3r^\prime} \, .
\label{eq:Purcell_extended_1}
\end{equation}
This expression of the normalized decay rate of an extended source accounts for the non-local interaction with the environment, as already put forward in Ref.~\cite{Lodahl2012}. The similarity with Eq.~(\ref{eq:Purcell_dipole}) can be made more explicit by introducing the projected CDOS~\cite{Caze2013}, defined as
\begin{equation}
\rho(\rg,\rg^\prime,\omega) = \frac{2 \omega}{\pi \, c^2} \, \mathrm{Im} \left [ \ug \cdot \Gg(\rg,\rg^\prime,\omega) \ug^\prime \right ] \, .
\label{eq:CDOS_def}
\end{equation}
The projected CDOS measures the connection between two unit dipoles $\ug$ and $\ug^\prime$ located at the points $\rg$ and $\rg^\prime$. More precisely, the CDOS counts the contribution of the electromagnetic modes connecting $\rg$ and $\rg^\prime$ at frequency $\omega$~\cite{Caze2013,Carminati-review,Sauvan2014}. It can also be understood as a measure of the intrinsic spatial coherence between the two points $\rg$ and $\rg^\prime$. We also note that for $\rg=\rg^\prime$ the projected CDOS coincides with the projected LDOS evaluated at point $\rg$. In terms of the CDOS, we find that
\begin{equation}
\frac{\Gamma(\rg_s,\omega)}{\Gamma_0(\omega)} = \frac{\int \eta^*(\rg,\rg_s) \, \eta(\rg^\prime,\rg_s) \, \rho(\rg,\rg^\prime,\omega) \, d^3r \, d^3r^\prime}{\int \eta^*(\rg,\rg_s) \, \eta(\rg^\prime,\rg_s) \, \rho_0(\rg,\rg^\prime,\omega) \, d^3r \, d^3r^\prime} \, ,
\label{eq:Purcell_extended}
\end{equation}
where $\rho_0(\rg,\rg^\prime,\omega)$ is the CDOS in free space. This expression of the normalized decay rate (or Purcell enhancement) generalizes Eq.~(\ref{eq:Purcell_dipole}) to extended emitters beyond the dipole approximation. Indeed, it can be seen immediately that by using $\eta(\rg,\rg_s)=p \, \delta(\rg-\rg_s)$ in Eq.~(\ref{eq:Purcell_extended}), which corresponds to a point dipole source with amplitude $p$ located at the point $\rg_s$, we recover Eq.~(\ref{eq:Purcell_dipole}).

\section{Extended sources in photonic cavities}

It is instructive to start with the simplest model of a {spatially coherent} extended source, made of two {mutually coherent} point dipoles with amplitude $\pg_1 = p_1 \ug_1$ and $\pg_2=p_2 \ug_2$, and located at positions $\rg_1$ and $\rg_2$, that is
\begin{equation}
\jg(\rg)=-i\omega \ug_1 p_1 \delta(\rg-\rg_1)  -i\omega \ug_2 p_2 \delta(\rg-\rg_2) \, .
\label{eq:source_simple}
\end{equation}
According to Eq.~(\ref{eq:Purcell_extended}), the spontaneous decay rate is
\begin{equation}
\Gamma \sim |p_1|^2 \rho_{11} + |p_2|^2 \rho_{22} + 2 \mathrm{Re}(p_1p_2^*) \rho_{12} \, ,
\end{equation}
up to a constant prefactor (independent of the environment) that we do not specify. Here $\rho_{ij} = 2\omega/(\pi c^2) \mathrm{Im} [\ug_i \cdot\Gg(\rg_i,\rg_j,\omega)  \ug_j ]$ is the LDOS for $i=j$ and the CDOS for $i \neq j$.
If the two dipoles have the same amplitude $|p|$ and a phase difference $\phi$, we simply have
\begin{equation}
\Gamma \sim |p|^2 [ \rho_{11} + \rho_{22} + 2 \rho_{12} \, \cos\phi] \, .
\label{eq:rate_CDOS}
\end{equation}
The mutual coherence between the two sources is essential for the appearance of the interference term involving the CDOS. This situation is different from that analyzed in Refs.~\cite{Canaguier2016,Canaguier2019,Carminati2015}, where spontaneous emission by two independent sources ({\it e.g.}, two fluorescent molecules) in a structured environment was considered. In this case, it was shown that intensity fluctuations are influenced by a cross term involving the CDOS.
In the situation considered here, interferences directly modify the spontaneous decay rate. The state of interference not only depends on the relative phase of the dipole sources, but also on the environment through the CDOS that can take positive or negative values. The increase or decrease of the spontaneous decay rate of an extended source due to interferences corresponds to superradiance and subradiance~\cite{Fauche2017,Lalanne2018}. In the following subsections, we use numerical simulations to illustrate the substantial role of the CDOS on the spontaneous decay rate of extended sources in photonic cavities. 

\subsection{High-$Q$ cavity}

Let us first consider a source immersed in a discrete set of high-$Q$ electromagnetic modes, whose resonances do not overlap spectrally. The Green's function in this case takes the following form~\cite{Carminati-review}
\begin{equation}
 \Gg(\rg,\rg^\prime,\omega) = \sum_m c^2 \frac{\eg_m(\rg) \otimes \eg_m^*(\rg^\prime)}{\omega_m^2-\omega^2-i\omega \gamma_m} \, ,
 \label{eq:Green_modes}
\end{equation}
where $\eg_m$ are eigenmodes of the vector Helmholtz equation in absence of losses, $\omega_m$ are eigenfrequencies, and $\gamma_m$ is a phenomenological mode damping rate. This expression can also be seen as low-loss limit ($\gamma_m \ll \omega_m$) of an expansion in quasi-normal modes (QNMs)~\cite{Sauvan2013}. For the CDOS we find that~\cite{Carminati-review,Sauvan2014}
\begin{equation}
\rho_{12} = \sum_m \frac{\gamma_m}{2\pi} \frac{\mathrm{Re} [(\ug_1\cdot\eg_m(\rg_1)) \, (\ug_2\cdot \eg_m^*(\rg_2))]}{(\omega-\omega_m)^2 + \gamma_m^2/4} \, .
 \label{eq:CDOS_modes}
\end{equation}
This expression shows explicitly that the CDOS accounts for the phase difference of each field mode $\eg_m$ between the points $\rg_1$ and $\rg_2$. In the particular case of a single-mode high-$Q$ cavity, the CDOS factorizes in the form $\rho_{12}^2 = \rho_{11} \rho_{22}$~\cite{Canaguier2019}. For a source made of two coherent dipoles located at positions $\rg_1$ and $\rg_2$ in a single mode cavity, we find from Eq.~(\ref{eq:rate_CDOS}) that the spontaneous decay rate is
\begin{equation}
\Gamma \sim |p|^2 [ \rho_{11} + \rho_{22} \pm 2 \sqrt{\rho_{11}\rho_{22}} \, \cos\phi] \, .
\label{eq:rate_CDOS_one_mode}
\end{equation}
The field mode in this case is a standing wave with real amplitude, and the plus or minus sign in the interference term corresponds to $\ug_1\cdot\eg_m(\rg_1)$ and $\ug_2\cdot\eg_m(\rg_2)$ having identical or opposite signs (which corresponds to a positive or negative CDOS $\rho_{12}$).

\begin{figure}[h]
     \begin{center}
     \includegraphics[width=9cm]{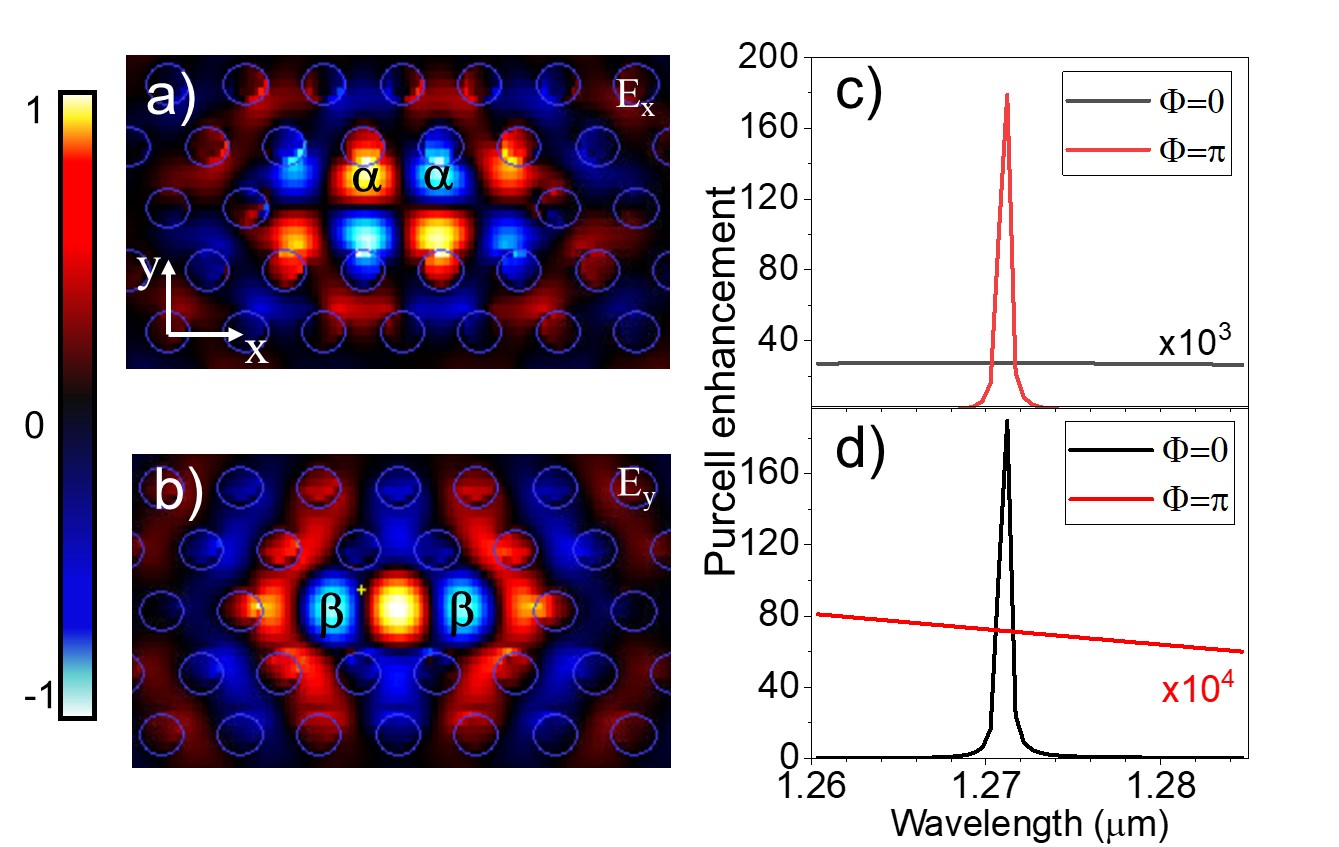}
     \end{center}    
     \caption{\label{fig:L3cavity} Emission by two dipoles in a L3 photonic crystal cavity on a slab. The slab has a refractive index $n=3.48$ and a thickness $L=200$~nm. The photonic crystal is made of holes with a diameter $d=180$~nm and a lattice constant $a=320$~nm. (a) and (b): Field amplitudes $E_x(\rg)= \mathrm{Re} [\xg \cdot \eg_m(\rg)]$ and $E_y(\rg)= \mathrm{Re} [\yg \cdot \eg_m(\rg)]$ of the cavity mode $\eg_m(\rg)$. The field is calculated in a plane coinciding with the middle of the slab. The points $\alpha$ and $\beta$ indicate the positions of the two emitting dipoles. The locations of the photonic crystal holes are also highlighted. (c) and (d): Spectrum of the normalized spontaneous decay rate $\Gamma/\Gamma_0$ (or Purcell enhancement) of two coherent dipoles with identical amplitudes, phase difference $\phi$, located at positions $\alpha$ and oriented along the $x$ direction (panel c), or located at positions $\beta$ and oriented along the $y$ direction (panel d).  Black line: $\phi=0$. Red line: $\phi=\pi$. The curves corresponding to subradiant emission have been multiplied by a factor of $10^3$ (panel c) and $10^4$ (panel d) for the sake of visibility.
     }
\end{figure}

In order to illustrate the relevance of Eq.~(\ref{eq:rate_CDOS_one_mode}) in a realistic situation, we consider a single mode contribution in the well-known L3 photonic crystal cavity~\cite{Noda2007}. The cavity is formed by a line of three missing holes in a triangular-lattice pattern of air holes on a slab
with refractive index $n=3.48$ and thickness $L=200$~nm. Calculated maps of the amplitudes $E_y(\rg)=\mathrm{Re}[\yg\cdot\eg_m(\rg)]$ and $E_x(\rg)=\mathrm{Re}[\xg\cdot\eg_m(\rg)]$ of the fundamental mode $m$, obtained from a FDTD simulation, are displayed in Fig.~\ref{fig:L3cavity}(a) and \ref{fig:L3cavity}(b).The positions of the photonic crystal holes, designing the L3 cavity, are indicated as circles. The fundamental mode $m$ is characterized by a LDOS with a Lorentzian lineshape (not shown), with a resonance wavelength $\lambda_m = 1270$~nm and a quality factor $Q_m=\omega_m/\gamma_m \simeq2000$. We focus on the emission by two coherent electric dipoles placed either at the points $\alpha$ indicated in Fig.~\ref{fig:L3cavity}(a), or at the points $\beta$ indicated in Fig.~\ref{fig:L3cavity}(b). We show in Fig.~\ref{fig:L3cavity}(c) [resp.~Fig.~\ref{fig:L3cavity}(d)] the Purcell enhancement $\Gamma/\Gamma_0$ for two dipoles with identical amplitude and phase difference $\phi$, and oriented along the $x$ direction [resp. $y$ direction]. The black curve corresponds to two dipoles in phase ($\phi=0$), and the red curve to two dipoles with opposite signs ($\phi=\pi$). The decay rate is obtained from a FDTD simulation, in which the emitted power is calculated from the net flux of the Poynting vector through a box containing the two dipoles, and having spectral linewidth much larger than the linewidth of the cavity mode.
For two dipoles placed at positions $\alpha$ [Fig.~\ref{fig:L3cavity}(c)], that are connected by a negative CDOS since the field amplitudes have opposite signs, we observe a subradiant emission for $\phi=0$ and perfect superradiance for $\phi=\pi$ (we have verified that the emission rate in this case is twice that obtained for a single dipole). For two dipoles placed at positions $\beta$ [Fig.~\ref{fig:L3cavity}(d)], connected by a positive CDOS, we observe superradiance for $\phi=0$ and subradiance for $\phi=\pi$, as expected. Note that the distances between the two points $\alpha$ and $\beta$ are $\Delta r=320$~nm and $\Delta r=480$~nm, respectively, that is comparable or larger than $\lambda/n$, with $\lambda$ the emission wavelength in vacuum. In bulk materials, super or subradiant effects would be observable only for $\Delta r \ll \lambda/n$. The simulations clearly confirm that the photonic mode imposes the constructive or destructive interference, through the sign of the CDOS, in agreement with Eq.~(\ref{eq:rate_CDOS_one_mode}).

\begin{figure}[h]
     \begin{center}
     \includegraphics[width=9cm]{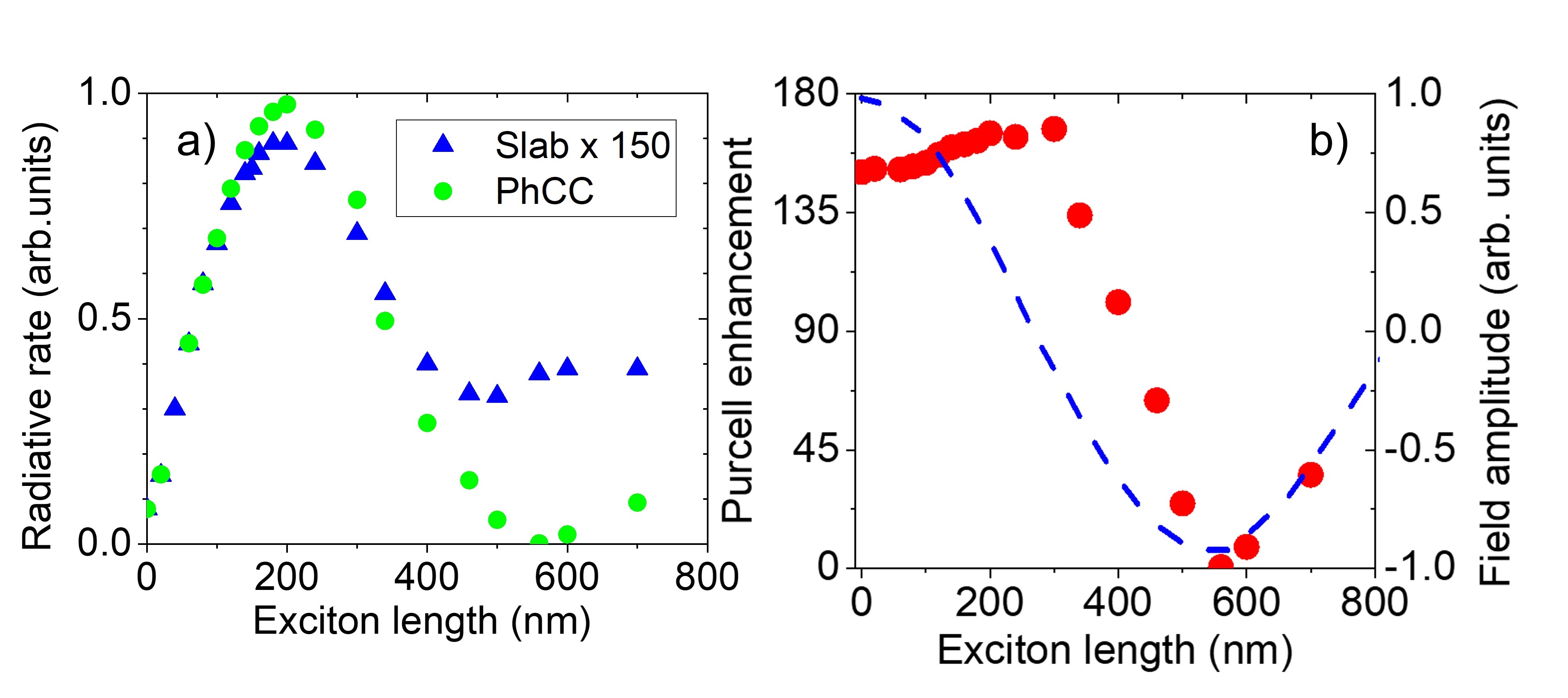}
     \end{center}    
     \caption{\label{fig:extendedL3}  (a) Comparison of the spontaneous decay rate of an emitting nanowire polarized along the $y$ direction and extended along the $x$ direction in the L3 cavity (green circles), and in a homogeneous medium with refractive index $n=3.48$ (blue triangles), versus the nanowire length $d$. In the case of the L3 cavity, the center of the nanowire coincides with the center of the cavity. (b) Purcell enhancement $\Gamma/\Gamma_0$ versus the emitting nanowire length (red dots). The field amplitude $E_y(\rg)= \mathrm{Re} [\yg \cdot \eg_m(\rg)]$ of the cavity mode at the extremities of the emitting nanowire is also indicated (blue dashed line). Up to a length $d \simeq 300$~nm, the center and the extremities of the nanowire are connected by a positive CDOS (the field amplitude is positive at the center and at the extremities). For $d \gtrsim 300$~nm, the center and the extremities of the emitting nanowire are connected by a negative CDOS, and destructive interferences reduce the spontaneous decay rate. The minimum is observed for $d \simeq 560$~nm.
}
\end{figure}

Next, in order to mimic the emission by a one-dimensional extended exciton (as found for example in a carbon nanotube or a quantum wire), {we consider a line source with length $d$ (hereafter denoted by emitting nanowire). In the FDTD simulation, the source model corresponding to the emission by a single exciton is a linear cluster of $N$ elementary and mutually coherent dipoles with amplitude $\pg/\sqrt{N}$ (as already pointed out the source is spatially coherent for a single exciton)}. In order to focus on the influence of the environment, we consider the situation where all elementary dipoles are in phase (but the approach allows one to use a more refined source model). The emitting nanowire is embedded in the L3 cavity, and we can expect very different radiative rates depending on its location and orientation. As a case study, we consider a nanowire oriented along the $x$ direction, polarized along the $y$ direction, and located at the center of the L3 cavity. The emission wavelength $\lambda=1270$~nm is on resonance with the cavity mode, and the nanowire can be seen as immersed in the field distribution shown in Fig.~\ref{fig:L3cavity}(b). We show in Fig.~\ref{fig:extendedL3}(a) the calculated radiative rate versus the nanowire length $d$ in the L3 cavity (green circles), and in a homogeneous medium with the same refractive index $n=3.48$ as the cavity slab (blue triangles). The radiative rate in the homogeneous medium increases with the exciton size, in agreement with the result already found in Ref.~\cite{Lodahl2012}, up to a maximum reached for a size $d=\lambda/2n$. Interestingly, a very similar trend is observed for the nanowire in the L3 cavity. 

To address this length dependence in more details, we show in Fig.~\ref{fig:extendedL3}(b) the Purcell enhancement $\Gamma/\Gamma_0$ in the L3 cavity (red dots) versus the emitting nanowire length. In the same figure we also display the field amplitude of the cavity mode at the extremities of the nanowire (blue dashed line). We see that the Purcell enhancement is almost constant for $d \leq 300$~nm, {\it i.e.} as long as the nanowire is inside the central hot spot of the field amplitude seen in Fig.~\ref{fig:L3cavity}(b). In this case, the CDOS remains positive over the emitting nanowire extent (the field amplitude of the mode does not change sign from the center to the extremities of the nanowire). Constructive interferences between the emissions from different parts of the nanowire lead to a substantial Purcell enhancement. However, for $d \gtrsim 300$~nm, the center and the extremities of the emitting nanowire are connected by a negative CDOS (the field amplitude of the mode is positive at the center of the nanowire and negative at the extremities, since they overlap with the negative spots indicated as $\beta$ in Fig.~\ref{fig:L3cavity}(b)). Destructive interferences reduce the spontaneous decay rate, leading to a minimum of the Purcell enhancement for $d \simeq 560$~nm. This subradiance of the emitting nanowire arises from the dominance of negative CDOS contributions, induced by the spatial structure of the underlying cavity mode.

\subsection{Coupled cavities with non-Lorentzian spectrum}

In addition to generating subradiance or superradiance, the CDOS contribution in the interferometric emission process may also influence the spectral lineshape. An interesting situation is that of two coupled photonic cavities generating a LDOS with a non-Lorentzian spectrum, which can be achieved by selecting two L3 cavities with unbalanced losses~\cite{Gurioli2020}. This situation goes beyond the high-$Q$ model used above, and the expansion (\ref{eq:CDOS_modes}) for the LDOS and CDOS has to be replaced by a more general expansion in QNMs (for an introduction to QNMs in photonics, see for example~\cite{Lalanne2018}). Considering that each cavity can be described by a single QNM, the Green's function takes the form~\cite{Sauvan2013,Sauvan2014}
\begin{equation}
 \Gg(\rg,\rg^\prime,\omega) \simeq  \frac{c^2}{2 \omega} \left [ \frac{\QEg_a(\rg) \otimes \QEg_a(\rg^\prime)}{\Qomega_a - \omega} + \frac{\QEg_b(\rg) \otimes \QEg_b(\rg^\prime)}{\Qomega_b - \omega} \right ]\, .
 \label{eq:Green_QNM}
\end{equation}
Here $\QEg_a$ and $\QEg_b$ are QNMs corresponding to cavity $a$ and $b$, respectively, with $\Qomega_a$ and $\Qomega_b$ the associated complex resonance frequencies. A major difference with Eq.~(\ref{eq:Green_modes}) is that the expression above is not limited to the low-loss (or high-$Q$) regime.

Coming back to the simple source model described by Eq.~(\ref{eq:source_simple}), with two coherent dipoles at positions $\rg_1$ and $\rg_2$, we can derive
the expressions of the LDOS and CDOS entering Eq.~(\ref{eq:rate_CDOS}). Defining $\omega_m=\mathrm{Re} \Qomega_m$ and $\gamma_m=-2\mathrm{Im} \Qomega_m$, with $m\in \{a,b\}$, the result can be cast in a form involving Fano profiles of the form
\begin{equation}
F(m,q,\omega) = \frac{\gamma_m/2}{q^2+1} \left [ \frac{(q^2-1)\gamma_m/2 + 2q(\omega-\omega_m)}{(\omega-\omega_m)^2 + \gamma_m^2/4} \right ] \, ,
\label{eq:Fano_general}
\end{equation}
where $q$ is a parameter defining the resonance lineshape~\cite{Gurioli2021}. Introducing the phases of the QNMs at the source points $\varphi_i^m=\arg[\ug_i\cdot\QEg_m(\rg_i)]$, with $i\in \{1,2\}$, and the parameters $q_i^m=\tan(\varphi_i^m)$, $q_{12}^m=[\tan(\varphi_1^m)+\tan(\varphi_2^m)]/2$, a simple calculation leads to
\begin{align}
&\rho_{11} \sim  F(a,q_1^a,\omega) + F(b,q_1^b,\omega) \label{eq:LDOS_Fano_1} \\
&\rho_{22} \sim  F(a,q_2^a,\omega) + F(b,q_2^b,\omega)  \label{eq:LDOS_Fano_2} \\
&\rho_{12} \sim  F(a,q_{12}^a,\omega) + F(b,q_{12}^b,\omega) \, .
\label{eq:CDOS_Fano}
\end{align}
We find that the spectrum of the LDOS and CDOS is a sum of Fano lineshapes, each arising from one QNM ($a$ and $b$), with $q$ parameters depending on the location of the source points ($1$ and $2$). These expressions reveal the complexity of the prediction of the spontaneous emission dynamics of an extended source in a structured environment, that depends on the LDOS and the CDOS, and on the spatial distribution of the phase of the QNMs. In particular, it is interesting to note that the spectrum of the decay rate cannot be predicted solely from the knowledge of the spectrum of the LDOS at each point. The relative phases $\varphi_i^m$ between different source points, imposed by the electromagnetic QNMs, play a crucial role on the lineshape of the emission spectrum.

\begin{figure}[h]
     \begin{center}
     \includegraphics[width=9cm]{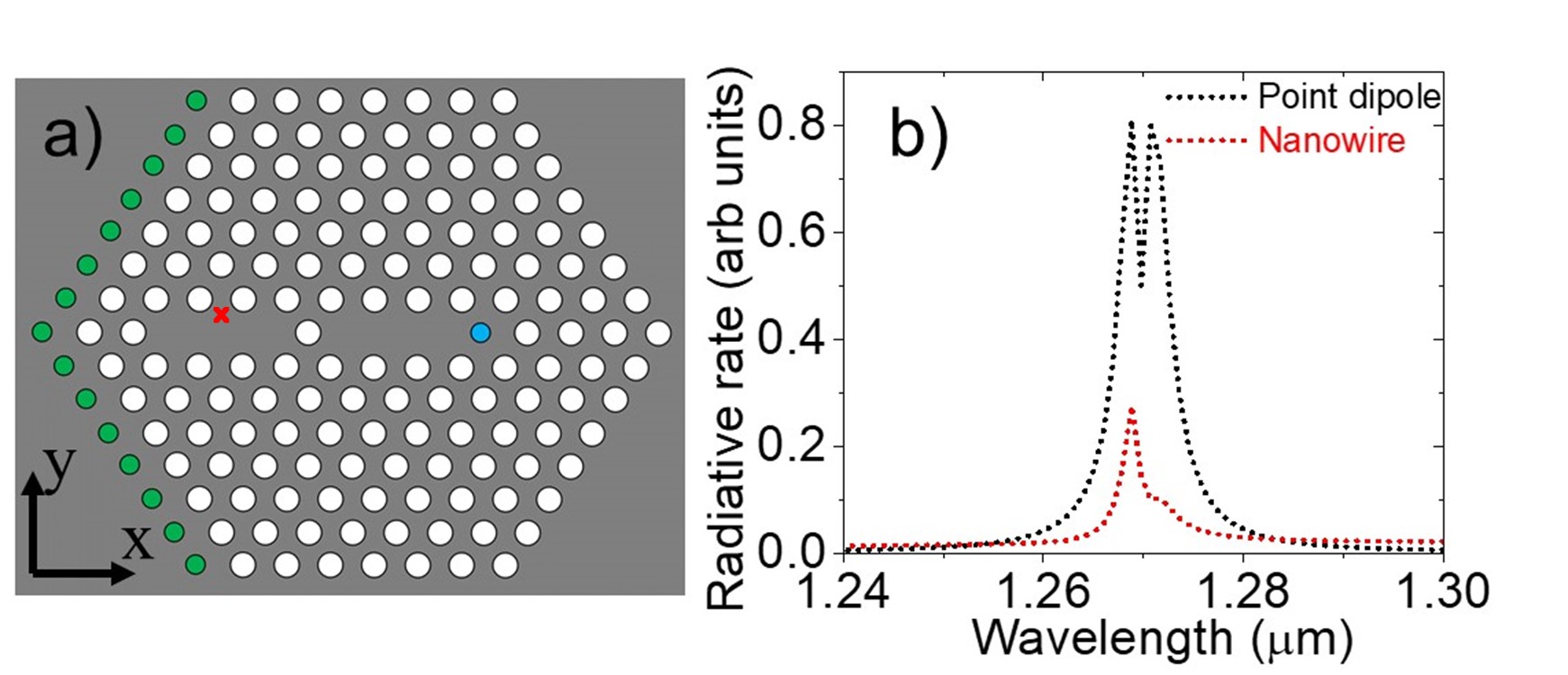}
     \end{center}    
     \caption{\label{fig:coupled_cavities} (a): Sketch of the system made of two coupled photonic-crystal cavities on a slab, with the same thickness and refractive index as in Fig.~\ref{fig:L3cavity}. The white holes have a diameter of 180~nm, the green holes of 140~nm and the blue hole of 150 nm. (b): Spectrum of the decay rate for a point dipole source (black dashed line) and an emitting nanowire (red dashed line). The red cross in the left cavity in panel (a) indicates the location of the center of the source, that coincides with one of the $\alpha$ points indicated in Fig.~\ref{fig:L3cavity}(a). Both sources are polarized along the $x$ direction. The emitting nanowire has a length $d=300$ nm and is oriented along the $x$ direction.
     }
\end{figure}

To illustrate this result, we have carried out numerical simulations in the system depicted in Fig.~\ref{fig:coupled_cavities}(a), made of two coupled photonic-crystal cavities with unbalanced losses. The holes with different colors have different radii, the green holes being designed to tailor the losses, and the pale blue hole being designed to reduce the frequency detuning between the two cavities. Unbalanced losses lead to highly non-Lorentzian LDOS and CDOS spectra (not shown), resulting from the superposition of two strongly asymmetric Fano contributions according to Eqs.~(\ref{eq:LDOS_Fano_1}-\ref{eq:CDOS_Fano}). As a result, we can expect the spectral profile of the spontaneous decay rate of a point dipole source to be very different from that of an extended exciton. This is clearly seen in Fig.~\ref{fig:coupled_cavities}(b), where we show the calculated spectrum of the spontaneous decay rate $\Gamma$ of a point dipole (black dashed line) and an emitting nanowire (red dashed line). Both sources are polarized in the $x$ direction, and centered on the position indicated by the red dot in Fig.~\ref{fig:coupled_cavities}(a). The nanowire has a length $d=300$ nm and is oriented along the $x$ direction. We observe that the CDOS contribution not only produces a subradiant effect reducing the decay rate to very small values, but also induces a substantial reshaping of the spectral profile. This example clearly demonstrates that the emission lineshape in the Purcell effect of an extended source cannot be predicted from the sole knowledge of the LDOS spectrum at all points. 

\section{Conclusion}

In summary, we have studied theoretically the change in the spontaneous decay rate (or Purcell effect) of an extended emitter in a structured environment, emphasizing the role of the CDOS in driving interferences in the emission process. Beyond the generation of large effective oscillator strengths, that had been put forward and exploited in previous studies, we have demonstrated that a structured CDOS (in space and/or spectrum) may produce subradiance or superradiance, and change substantially the emission spectrum. These results, based on a simple but robust theory, have been highlighted by numerical simulations in realistic photonic-crystal cavity geometries. The study clearly shows that a knowledge of the LDOS spectrum at all points does not permit to predict the change in the decay rate, neither in value nor in lineshape. It also emphasizes the importance of the engineering of both the LDOS and the CDOS for the control of spontaneous emission by extended sources.

\section*{Acknowledgments}
We thank P. Lalanne and K. Vynck for helpful discussions at the initial stage of this study. M.G. acknowledges a visiting fellowship from Institut Langevin and ESPCI Paris - PSL where part of this work was done.

\section*{Funding}
This work was supported by LABEX WIFI (Laboratory of Excellence within the French Program Investments for the Future) under references ANR-10- LABX-24 and ANR-10-IDEX-0001-02 PSL*.



\begin{thebibliography}{40}

\bibitem{cavityQED}
{\it Cavity Quantum Electrodynamics}, edited by P. Berman (Academic Press, 1994);
S. Haroche and J.M. Raimond, {\it Exploring the Quantum: Atoms, Cavities, and Photons} (Oxford University Press, 2006).

\bibitem{Shields2007}
A.J. Shields, ``Semiconductor quantum light sources,'' Nature Phot. {\bf 1}, 215-223 (2007).

\bibitem{Stockman-review}
M.I. Stockman, ``Nanoplasmonics: past, present, and glimpse into future,'' Opt. Express {\bf 19}, 22029-22106 (2011).

\bibitem{Novotny-book}
L. Novotny and B. Hecht, {\it Principles of Nano-Optics} (Cambridge University Press, 2006).

\bibitem{Carminati-review}
R. Carminati, A. Caz\'e, D. Cao, F. Peragut, V. Krachmalnicoff, R. Pierrat, and Y. De Wilde, ``Electromagnetic density of states in complex plasmonic systems,'' Surf. Sci. Rep. {\bf 70}, 1-41 (2015).

\bibitem{Purcell1946}
E. M. Purcell, ``Spontaneous emission probabilities at radio frequencies,'' Phys. Rev. {\bf 69}, 681 (1946).

\bibitem{Andreani1999}
L.C. Andreani, G. Panzarini and J.M. G\'erard, ``Strong-coupling regime for quantum boxes in pillar microcavities: Theory,'' Phys. Rev. B {\bf 60}, 13276 (1999).

\bibitem{Benz2016}
F. Benz, M.K. Schmidt, A. Dreismann, R. Chikkaraddy, Y. Zhang, A. Demetriadou, C. Carnegie, H. Ohadi, B. de Nijs, R. Esteban, J. Aizpurua and J.J. Baumberg, ``Single-molecule optomechanics in picocavities,'' Science {\bf 354}, 726-729 (2016).

\bibitem{Lodahl2012}
S. Stobbe, P. T. Kristensen, J. E. Mortensen, J. M. Hvam, J. M\o rk and P. Lodahl, ``Spontaneous emission from large quantum dots in nanostructures: Exciton-photon interaction beyond the dipole approximation,'' Phys. Rev. B {\bf 86}, 085304 (2012).

\bibitem{Kazimierczuk2014}
T. Kazimierczuk, D. Fr\"ohlich, S. Scheel, H. Stolz and M. Bayer, ``Giant Rydberg Excitons in Cuprous Oxide,'' Nature {\bf 514}, 343-347 (2014).

\bibitem{Steinhauer2020}
S. Steinhauer, M.A.M. Versteegh, S. Gyger, A. W. Elshaari, B. Kunert, A. Mysyrowicz and V. Zwiller, ``Rydberg excitons in Cu2O microcrystals grown on a silicon platform,'' Commun. Materials {\bf 1}, 11 (2020).

\bibitem{Haroche2013}
S. Haroche, ``Controlling photons in a box and exploring the quantum to classical boundary,'' Rev. Mod. Phys. {\bf 85}, 1083 (2013).

\bibitem{Abajo2010}
F. J. Garc\'\i a de Abajo, ``Optical excitations in electron microscopy,'' Rev. Mod. Phys. {\bf 82}, 209 (2010).

\bibitem{Kimble2008}
H. J. Kimble, ``The quantum internet,'' Nature (London) {\bf 453}, 1023-1030 (2008).

\bibitem{Pryde2019}
S. Slussarenko and  G.J. Pryde, ``Photonic quantum information processing: A concise review,'' Appl. Phys. Reviews {\bf 6}, 041303 (2019).

\bibitem{Pustovit2009}
V. N. Pustovit and T. V. Shahbazyan, ``Cooperative emission of light by an ensemble of dipoles near a metal nanoparticle: The plasmonic Dicke effect,'' Phys. Rev. Lett. {\bf 102}, 077401 (2009).

\bibitem{Oppel2014}
S. Oppel, R. Wiegner, G. S. Agarwal, and J. von Zanthier, ``Directional Superradiant Emission from Statistically Independent Incoherent Nonclassical and Classical Sources,'' Phys. Rev. Lett. {\bf 113}, 263606 (2014).

\bibitem{Chang2017}
M.T. Manzoni, D.E. Chang and J.S. Douglas, ``Simulating quantum light propagation through atomic ensembles using matrix product states,'' Nat. Commun. {\bf 8}, 1743 (2017).

\bibitem{Feist2021}
J. Feist, A. I. Fern\'andez-Dom\'\i nguez and F.J. Garc\'\i a-Vidal, ``Macroscopic QED for quantum nanophotonics: emitter-centered modes as a minimal basis for multiemitter problems,'' Nanophotonics {\bf 10}, 477-489 (2021).

\bibitem{Caze2013}
A. Caz\'e, R. Pierrat, and R. Carminati, ``Spatial Coherence in Complex Photonic and Plasmonic Systems,'' Phys. Rev. Lett. {\bf 110}, 063903 (2013).

\bibitem{Canaguier2016}
A. Canaguier-Durand and R. Carminati, ``Quantum coherence of light emitted by two single-photon sources in a structured environment,'' Phys. Rev. A {\bf 93}, 033836 (2016).

\bibitem{Canaguier2019}
A. Canaguier-Durand, R. Pierrat and R. Carminati, ``Cross density of states and mode connectivity: Probing wave localization in complex media,'' Phys. Rev. A {\bf 99}, 013835 (2019).

\bibitem{Joulain2003}
K. Joulain, R. Carminati, J.P. Mulet and J.-J. Greffet, ``Definition and measurement of the local density of electromagnetic states close to an interface,'' Phys. Rev. B {\bf 68}, 245405 (2003).

\bibitem{Aigouy2014}
L. Aigouy, A. Caz\'e, P. Gredin, M. Mortier and R. Carminati, ``Mapping and quantifying electric and magnetic dipole luminescence at the nanoscale,'' Phys. Rev. Lett. {\bf 113}, 076101 (2014).

\bibitem{Agarwal1975}
G. S. Agarwal, ``Quantum electrodynamics in the presence of dielectrics and conductors. I. Electromagnetic-field response functions and black-body fluctuations in finite geometries,'' Phys. Rev. A {\bf 11}, 230 (1975).

\bibitem{Sipe1984}
J. M. Wylie and J. E. Sipe, ``Quantum electrodynamics near an interface,'' Phys. Rev. A {\bf 30}, 1185 (1984).

\bibitem{Bouchet2019}
D. Bouchet and R. Carminati, ``Quantum dipole emitters in structured environments: a scattering approach: tutorial,'' J. Opt. Soc. Am. A {\bf 36}, 186-195 (2019).

\bibitem{Sauvan2013}
C. Sauvan, J.P. Hugonin, I. S. Maksymov and P. Lalanne, ``Theory of the Spontaneous Optical Emission of Nanosize Photonic and Plasmon Resonators,'' Phys. Rev. Lett. {\bf 110}, 237401 (2013).

\bibitem{Sauvan2014}
C. Sauvan, J.P. Hugonin, R. Carminati and P. Lalanne, ``Modal representation of spatial coherence in dissipative and resonant photonic systems,'' Phys. Rev. A {\bf 89}, 043825 (2014).

\bibitem{Carminati2015}
R. Carminati, G. Cwilich, L.S. Froufe P\'erez and J.J. S\'aenz, ``Speckle fluctuations resolve the interdistance between incoherent point sources in complex media,'' Phys. Rev. A {\bf 91}, 023807 (2015).

\bibitem{Fauche2017}
P. Fauch\'e, S.G. Kosionis and P. Lalanne, ``Collective scattering in hybrid nanostructures with many atomic oscillators coupled to an electromagnetic resonance,'' Phys. Rev. B {\bf 95}, 195418 (2017).

\bibitem{Lalanne2018}
P. Lalanne, W. Yan, K. Vynck, C. Sauvan and J.P. Hugonin, ``Light Interaction with Photonic and Plasmonic Resonances,'' Laser Phot. Rev. {\bf 12}, 1700113 (2018).

\bibitem{Noda2007}
S. Noda, M. Fujita and T. Asano, ``Spontaneous-emission control by photonic crystals and nanocavities,'' Nat. Phot. {\bf 1}, 449-458 (2007).

\bibitem{Gurioli2020}
D. Pellegrino, D. Balestri, N. Granchi, M. Ciardi, F. Intonti, F. Pagliano, A.Y. Silov, F. W. Otten, T. Wu, K. Vynck, P. Lalanne, A. Fiore and M. Gurioli, ``Non-Lorentzian Local Density of States in Coupled Photonic Crystal Cavities Probed by Near- and Far-Field Emission,'' Phys. Rev. Lett. {\bf 124}, 123902 (2020).

\bibitem{Gurioli2021}
M. Gurioli, ``Vacuum Field Photonic Trap for Excitons,'' Adv. Quantum Technol. {\bf 4}, 2100046 (2021).

\end{thebibliography}
\end{document}